\documentclass[aps,prl,twocolumn,superscriptaddress, noeprint]{revtex4-2}
\usepackage{graphicx}
\usepackage{chemformula}
\usepackage{siunitx}
\usepackage{gensymb}
\usepackage{epstopdf}

\begin{document}
	
	\title{Quantum Dot-Based Parametric Amplifiers}
	
	\author{Laurence Cochrane}
	\email{olc22@cam.ac.uk}
	\affiliation{Nanoscience Centre, Department of Engineering, University of Cambridge, Cambridge CB3 0FF, UK}
	\affiliation{Quantum Motion Technologies, Windsor House, Cornwall Road, Harrogate HG1 2PW, UK}
	\author{Theodor Lundberg}
	\affiliation{Cavendish Laboratory, University of Cambridge, J.J. Thomson Avenue, Cambridge CB3 0HE, UK}
	\affiliation{Hitachi Cambridge Laboratory, J.J. Thomson Avenue, Cambridge CB3 0HE, UK}
	\author{David J. Ibberson}
	\affiliation{Quantum Motion Technologies, Windsor House, Cornwall Road, Harrogate HG1 2PW, UK}
	\author{Lisa Ibberson}
	\affiliation{Hitachi Cambridge Laboratory, J.J. Thomson Avenue, Cambridge CB3 0HE, UK}
	\author{Louis Hutin}
	\affiliation{CEA/LETI-MINATEC, CEA-Grenoble, 38000 Grenoble, France}
	\author{Benoit Bertrand}
	\affiliation{CEA/LETI-MINATEC, CEA-Grenoble, 38000 Grenoble, France}
	\author{Nadia Stelmashenko}
	\affiliation{Department of Materials Science and Metallurgy, University of Cambridge, 27 Charles Babbage Road, Cambridge CB3 0FS, United Kingdom}
	\author{Jason W. A. Robinson}
	\affiliation{Department of Materials Science and Metallurgy, University of Cambridge, 27 Charles Babbage Road, Cambridge CB3 0FS, United Kingdom}
	\author{Maud Vinet}
	\affiliation{CEA/LETI-MINATEC, CEA-Grenoble, 38000 Grenoble, France}
	\author{Ashwin A. Seshia}
	\affiliation{Nanoscience Centre, Department of Engineering, University of Cambridge, Cambridge CB3 0FF, UK}
	\author{M. Fernando Gonzalez-Zalba}
	\affiliation{Quantum Motion Technologies, Windsor House, Cornwall Road, Harrogate HG1 2PW, UK}

	\date{\today}
	
	\begin{abstract}
		Josephson parametric amplifiers (JPAs) approaching quantum-limited noise performance have been instrumental in enabling high fidelity readout of superconducting qubits and, recently, semiconductor quantum dots (QDs). We propose that the quantum capacitance arising in electronic two-level systems (the dual of Josephson inductance) can provide an alternative dissipation-less non-linear element for parametric amplification. We experimentally demonstrate phase-sensitive parametric amplification using a QD-reservoir electron transition in a CMOS nanowire split-gate transistor embedded in a 1.8~GHz superconducting lumped-element microwave cavity, achieving parametric gains of -3 to +3~dB, limited by Sisyphus dissipation. Using a semi-classical model, we find an optimised design within current technological capabilities could achieve gains and bandwidths comparable to JPAs, while providing complementary specifications with respect to integration in semiconductor platforms or operation at higher magnetic fields.
	\end{abstract}
	
	\maketitle
	
	Fast, high-fidelity qubit readout will be essential for fault-tolerant quantum computation~\cite{fowlerSurfaceCodesPractical2012}. The low energy scales typical of solid-state qubit implementations and the requirement to measure faster than the qubit coherence time $T_2$ necessitate high sensitivity, ultra-low noise experimental techniques. In the case of dispersive readout, a quantum non-demolition method to infer the qubit state from a state-dependent shift in the frequency of a non-resonantly coupled microwave cavity (the standard for superconducting qubits~\cite{krantzQuantumEngineerGuide2019, walterRapidHighFidelitySingleShot2017, aruteQuantumSupremacyUsing2019} and of increasing prevalence in gate-based implementations for semiconductor qubits~\cite{zhengRapidGatebasedSpin2019,westGatebasedSingleshotReadout2019, scarlinoAllMicrowaveControlDispersive2019, miCoherentSpinPhoton2018}), the cavity may be populated by less than a single photon on average: amplification is required before quadrature demodulation, hence the noise added by the first stage limits the minimum measurement time to discern the qubit state. To improve upon state-of-the-art cryogenic amplifiers (with typical noise temperatures of a few Kelvin~\cite{weinrebDesignCryogenicSiGe2007}) and approach quantum-limited noise performance ~\cite{clerkIntroductionQuantumNoise2010}, researchers have turned their attention to parametric amplifiers~\cite{royIntroductionParametricAmplification2016,aumentadoSuperconductingParametricAmplifiers2020, miStrongCouplingSingle2017, mehrpooCryogenicCMOSParametric2020}. 
  	
\begin{figure}
		\includegraphics{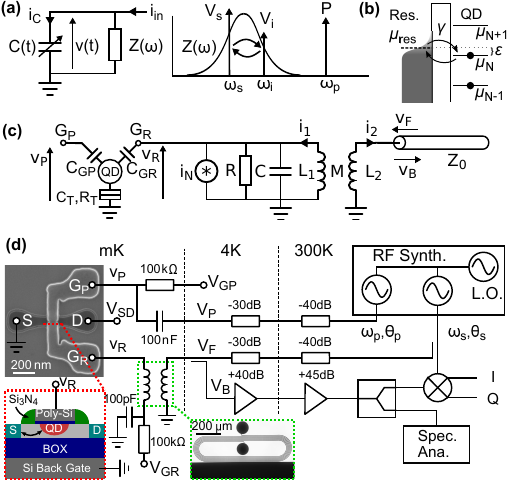}
		\caption{Concept of the Quantum Dot Parametric Amplifier (QDPA) and experimental implementation. (a) Simplest manifestation of a parametric amplifier as a pumped variable reactance $C$ and band-limited load $Z$, and the mechanism of energy exchange between signal and idler modes. (b) Origin of quantum capacitance in a quantum dot tunnel coupled to a reservoir. (c) Realisation of a parametric amplifier using a two-gate quantum dot as a variable capacitance and an $LC$ resonator inductively coupled to a feed line as the load. (d) Experimental implementation of (c) using a split-gate CMOS nanowire field-effect transistor and a superconducting spiral inductor. Resistor/capacitor bias tees are implemented at mK to set the gate potentials and deliver RF signals generated by a synthesiser, with a measurement chain comprised of cryogenic and room temperature amplification, I/Q demodulation and a spectrum analyser.
		\label{fig:fig1}}
	\end{figure}

	Parametric amplification is based on energy conversion between the `signal' mode (frequency $\omega_s$) and an auxiliary `idler' mode (at $\omega_i$), mediated by pumping a non-linear or variable reactive element at $\omega_p = \omega_s +\omega_i$ (Fig.~\ref{fig:fig1}(a)). Replacing the varactor-based designs of the 1960s, current superconducting implementations utilising the non-linear inductance of Josephson junctions (JJs), including resonant flux-pumped \cite{yamamotoFluxdrivenJosephsonParametric2008, muckRadiofrequencyAmplifiersBased2010} or current-pumped ~\cite{vijayInvitedReviewArticle2009} Josephson Parametric Amplifiers (JPAs) and Travelling Wave Parametric Amplifiers (TWPAs) ~\cite{obrienResonantPhaseMatching2014}, have proved vital making superconducting qubit-based architectures practicable by enabling single shot readout with high fidelity~\cite{walterRapidHighFidelitySingleShot2017,aumentadoSuperconductingParametricAmplifiers2020}. More recently, JPAs have produced significant performance enhancements in the readout of semiconductor quantum dots~\cite{stehlikFastChargeSensing2015, schaalFastGateBasedReadout2020, schuppSensitiveRadiofrequencyReadout2020}, though the magnetic shielding required presents challenges for closer integration with spin qubits. The limited dynamic range of JJ-based designs has encouraged research into using alternative non-linearities for parametric amplification, such as kinetic inductance~\cite{malnouThreeWaveMixingKinetic2021, malnouPerformanceKineticInductanceTravelingWave2021, parkerNearidealDegenerateParametric2021}.
	
	Beyond JJs, another non-linear reactive element that manifests in low-dimensional systems is quantum capacitance: rapid variations of the number of states with respect to the Fermi energy can generate an additional differential capacitance that is dual of the Josephson inductance~\cite{ashooriSingleelectronCapacitanceSpectroscopy1992, buttikerMesoscopicCapacitors1993}. The applications of quantum capacitance in artificial two-level systems, such as the Cooper-pair box or semiconductor quantum dots (QDs), have been primarily oriented towards compact, high-fidelity quantum state readout~\cite{peterssonChargeSpinState2010, pakkiamSingleShotSingleGateRf2018, borjansSpinDigitizerHighfidelity2021} and low-temperature thermometry \cite{ahmedPrimaryThermometrySingle2018, chawnerNongalvanicCalibrationOperation2021}. However, since tunnelling can be designed to be adiabatic, quantum capacitance devices may provide an alternative dissipation-less element for parametric amplification~\cite{suhParametricAmplificationBackAction2010}. This relatively unexplored area of research, particularly when focused on semiconductor-based devices, could provide amplifiers with complementary technological specifications to JJ-based parametric amplifiers, such as integration with semiconductor-based quantum computing circuits, resilience against magnetic fields, and compactness, especially if high-kinetic inductance materials are used for the resonant circuit \cite{samkharadzeHighKineticInductanceSuperconductingNanowire2016, bassetHighKineticInductance2019}. 
	
	In this Letter, we propose and demonstrate parametric amplification using the quantum capacitance of a silicon QD embedded in a lumped-element microwave cavity. The differential capacitance seen from the QD gate due to cyclic electron tunnelling between a QD energy level $N$ at electrochemical potential $\mu_N$ and a thermally broadened reservoir with a Fermi level $\mu_\mathrm{res}$ (Fig.~\ref{fig:fig1}(b)) depends on the QD excess electron occupancy probability, $P(\varepsilon/k_\mathrm{B}T)$, where the detuning $\varepsilon=\mu_N-\mu_\mathrm{res}$ \cite{ahmedPrimaryThermometrySingle2018}. At excitation frequencies $f$ below the QD-reservoir tunnel rate $\gamma$ and for temperatures $k_\mathrm{B}T\gg h\gamma$, tunnelling occurs adiabatically as $P$ tracks the thermal population, resulting in a dissipationless quantum capacitance
	\begin{equation}
		C_\mathrm{Q} = (e\alpha_\mathrm{R})^2\frac{\partial P}{\partial \varepsilon} = \frac{(e\alpha_\mathrm{R})^2}{4k_\mathrm{B}T}\frac{1}{\cosh^2(\varepsilon/2k_\mathrm{B}T)}
		\label{C_Q_thermal}
	\end{equation}
	where $\alpha_\mathrm{i}$ is the QD gate lever-arm $C_\mathrm{Gi}/(C_\mathrm{GR}+C_\mathrm{GP}+C_\mathrm{T})$ (Fig.~\ref{fig:fig1}(c)). A device with two gates ($\mathrm{G_R}$ and  $\mathrm{G_P}$) allows control of the detuning using either gate potential $v_\mathrm{R,P}$: $\varepsilon=-e\alpha_\mathrm{R}\Delta v_\mathrm{R}-e\alpha_\mathrm{P}\Delta v_\mathrm{P}$, where $\Delta v_\mathrm{R,P}=v_\mathrm{R,P}-V_\mathrm{R,P}^0$ and $(V_\mathrm{R}^0,V_\mathrm{P}^0)$ are gate voltage offsets that give $\mu_N=\mu_\mathrm{res}$. The device can then operate as a variable capacitor (as viewed from $\mathrm{G_R}$), tunable by the voltage applied to gate $\mathrm{G_P}$. To form a `Quantum Dot Parametric Amplifier' (QDPA), gate $\mathrm{G_R}$ is connected to a parallel $LC$ resonator while an excitation $v_p=V_p \cos(\omega_pt+\theta_p)$ applied to $\mathrm{G_P}$ pumps the quantum capacitance with amplitude $\Delta C_\mathrm{GR} = e\alpha_\mathrm{P}V_p\partial C_Q/\partial\varepsilon$ (Fig.~\ref{fig:fig1}(c)). With the resonator inductively coupled to a transmission line, the amplifier can operate in reflection, excited by a forward wave $v_f=V_f\cos(\omega_st+\theta_s)$ (the input signal) and emitting a backward wave $v_b=V_{bs}\cos(\omega_st+\theta_{bs})+V_{bi}\cos(\omega_it+\theta_{bi})$.
	
	In our experimental realisation (Fig.~\ref{fig:fig1}(d)) we use separate silicon and superconducting chips for the QD device and resonator respectively~\cite{ibbersonLargeDispersiveInteraction2021}. The QD is formed in a split-gate, fully-depleted nanowire field-effect transistor~\cite{lundbergSpinQuintetSilicon2020} with a 70~nm channel width, 60~nm gate length and a 40~nm split-gate separation, fabricated in a complementary metal-oxide-semiconductor (CMOS) silicon-on-insulator (SOI) process with an SOI thickness of 7~nm and buried oxide (BOX) thickness of 145~nm. For applied potentials $V_\mathrm{R}$ above a certain threshold, electrons accumulate in the corner of the square nanowire cross-section directly under $\mathrm{G_R}$ (see the red dashed line on the scanning electron micrograph and associated cross sectional sketch). Tunnel barriers between the QD and the source and drain contacts appear due to the ungated silicon region under the long (34~nm) \ch{Si_3N_4} spacers either side of the gate. The other gate of the split, $\mathrm{G_P}$, serves as the side gate for pumping. On the superconducting chip, a spiral inductor $L_1=\SI{47}{nH}$ is fabricated adjacent to a $Z_0=\SI{50}{\ohm}$ microstrip waveguide by optical lithography of an 80~nm thick \ch{NbN} film sputter-deposited on sapphire (see the optical micrograph inset in Fig.~\ref{fig:fig1}(d)). The parasitic capacitance $C$ of the inductor, wire bonds and bond pads completes the tank resonator. The inductor geometry and separation to the transmission line are designed to maximise the mutual inductance $M$ and exceed critical coupling of the resonator to the input transmission line. The wrap-around gate design combined with the high impedance of the resonator ($Z_r=\SI{540}{\ohm}$) enables achieving large QD-resonator coupling strengths~\cite{ibbersonLargeDispersiveInteraction2021}, a critical requirement to attain sufficient pump-driven capacitance modulation for parametric amplification.    
	
\begin{figure}
		\includegraphics{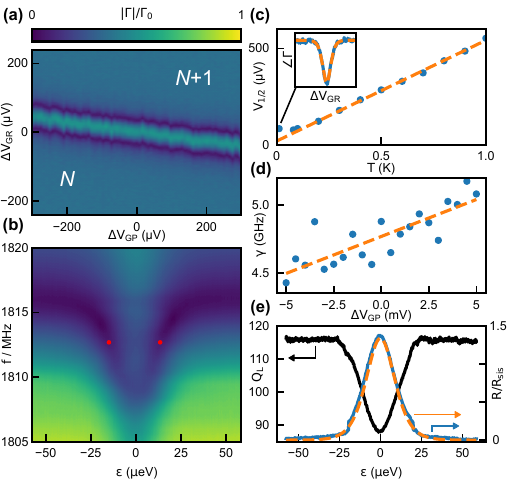}
		\caption{Characterisation of dot to reservoir transition and resonator properties from reflectometry data. (a) Charge stability diagram of normalised reflected magnitude as a function of gate potentials. (b) Resonator frequency shift as a function of detuning, with point of maximum $\partial f/\partial \varepsilon$ highlighted in red. (c) Lineshape analysis of transition FWHM $V_{1/2}$ as a function of temperature; the gate lever arm $\alpha_\mathrm{R}$ is determined from the gradient. (d) Dependence of lineshape on gate potential allowing calculation of the tunnel rate from the FWHM $\varepsilon_{1/2} = 2h\gamma$. (e) Variation of loaded Q factor with detuning and corresponding Sisyphus resistance, allowing fitting of the electron temperature.
		\label{fig:fig2}}
	\end{figure}

	We characterise the resonator and small-signal properties of a dot-to-reservoir (DTR) transition by probing the non-pumped circuit in reflectometry (Fig.~\ref{fig:fig2}(a)), measuring the change in reflection coefficient $\Gamma_0=V_b/V_f$ across the charge degeneracy point. From the unperturbed resonator frequency $f_0=1816$~MHz and maximum frequency shift $\Delta f=5.1$~MHz (Fig.~\ref{fig:fig2}(b)) we estimate a parasitic capacitance $C=\SI{163}{fF}$ and maximum quantum capacitance of \SI{0.93}{fF}. Fitting the resonance using a complex external Q-factor~\cite{kudraHighQualityThreedimensional2020} (see supplementary material), away from the charge transition we find a resonator-waveguide coupling coefficient $\beta=Q_i/Q_e=1.2$ and an internal Q-factor $Q_i=283$, modelled by a lumped element parallel resistance of $R=\SI{152}{\kilo\ohm}$. A lineshape analysis of the full-width half-maximum of the quantum capacitance signal as the temperature is varied (Fig.~\ref{fig:fig2}(c), $V_{1/2}=4\ln(\sqrt{2}+1)k_\mathrm{B}T/(e\alpha_\mathrm{R})$ from eq. \ref{C_Q_thermal}) gives $\alpha_\mathrm{R}=0.58$, while from the gradient of DTR transition line in the charge stability diagram  as the DC gate potentials $V_\mathrm{GR}$ and $V_\mathrm{GP}$ are varied (Fig.~\ref{fig:fig2}(a)) we find the ratio of lever arms $\alpha_\mathrm{R}/\alpha_\mathrm{P}=7.1$. Below 100~mK, the lineshape remains constant with temperature but becomes dependent on the potential applied to gate $\mathrm{G_P}$. This suggests that, at the operating temperature of 12~mK, the transition is lifetime broadened ($k_\mathrm{B}T<h\gamma$), with the tunnel barrier shape tuneable by the potential $V_\mathrm{GP}$ (see Fig.~\ref{fig:fig2}(d)). In this regime, the quantum capacitance is given by \cite{houseRadioFrequencyMeasurements2015} 
	\begin{equation}
		C_Q = \frac{(e\alpha_\mathrm{R})^2}{\pi}\frac{h\gamma}{(h\gamma)^2 + \varepsilon^2}
	\end{equation}
	and by fitting the FWHM $\varepsilon_{1/2} = 2h\gamma$, we find a tunnel rate at the nominal operating point $(V_\mathrm{R}^0, V_\mathrm{P}^0)$ used in subsequent experiments of 4.77~GHz, equating to a consistent value of $C_Q = \SI{0.87}{fF}$, in close agreement with the extracted quantum capacitance above. As the tunnel rate is comparable to $f_0$, a non-adiabatic tunnelling contribution is expected, leading to losses represented by a Sisyphus resistance of~\cite{perssonExcessDissipationSingleElectron2010, gonzalez-zalbaProbingLimitsGatebased2015}:
	\begin{equation}
		R_\mathrm{Sis} = \frac{2k_\mathrm{B}T}{\pi(e\alpha_\mathrm{R})^2} \frac{1 + \gamma^2/f^2}{\gamma} \cosh^2\left(\frac{\varepsilon}{2k_\mathrm{B}T}\right)
	\end{equation}
	This is in agreement with the decrease in loaded Q factor of the resonator observed at zero detuning, from which we estimate the minimum Sisyphus resistance to be \SI{110}{\kilo\ohm} and can fit an electron temperature of 67~mK (Fig.~\ref{fig:fig2}(e)). This source of dissipation is disadvantageous for parametric amplification, increasing the necessary pump power and inducing cyclostationary noise arising from stochastic electron tunnelling~\cite{gonzalez-zalbaProbingLimitsGatebased2015}. Though in our experiment large $\Delta f$ was prioritised in selection of the charge transition, adiabatic tunnelling can be engineered through shaping the tunnel barrier to achieve $\gamma \gg f$, or by exploiting quantum capacitance arising from the curvature of energy bands near avoided crossings, such as in a double quantum dot~\cite{mizutaQuantumTunnelingCapacitance2017}.
	
\begin{figure}
		\includegraphics{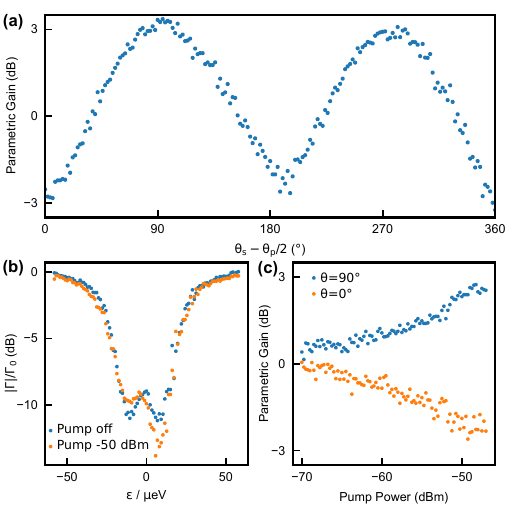}
		\caption{Evidence of parametric amplification. (a) Variation of reflected amplitude as a function of signal phase at a pump power of -50 dBm, with respect to the non-pumped circuit (b) Dependence of reflected amplitude with detuning at a signal frequency $\omega_s = \SI{1813.5}{MHz}$ with the pump off (blue) and on (orange) at -50 dBm. (c) Parametric gain as a function of pump amplitude for phases $0\degree$ (orange) and $90\degree$ (blue).
		\label{fig:fig3}}
	\end{figure}

	Having characterised the properties of resonator and the QD as a variable capacitor, we operate the circuit as a degenerate parametric amplifier by delivering phase-locked pump and probe signals to the device at $\omega_p = 2\omega_s$. To avoid saturation effects, the input signal power $V_f^2/2Z_0$ is kept at the low level of -135~dBm, while the pump power is in the range -70 to -47 dBm. The efficiency and dynamic range of the pump are optimised by selecting the detuning bias at the point of maximum $\partial{f}/\partial\varepsilon$ and hence maximum $\partial{C_Q}/\partial V_P$, corresponding to a resonant frequency of 1813.5~MHz. By sweeping the phase of the signal $\theta_s$ relative to the phase of the pump, we observe a $180\degree$-periodic variation in the reflected magnitude relative to the non-pumped system, with a maximum amplification and de-amplification of $\pm3$~dB (Fig. \ref{fig:fig3}(a)). Sweeping the detuning across the transition while maintaining  $\omega_s = \omega_p/2 = \SI{1813.5}{MHz}$ (Fig. \ref{fig:fig3}(b), equivalent to a horizontal section in Fig \ref{fig:fig2}(b)) results in amplification on one side and de-amplification on the other, consistent with the change in sign of $\partial{C_Q}/\partial V_P$ giving a $180\degree$ change in pump phase $\theta_p$. Finally, we sweep the pump amplitude while maintaining the relative phases at the points of maximum amplification and de-amplification ($\theta_s = 90$ and $0\degree$ respectively). We find the gain saturates at $\SI{\pm 3}{dBm}$ (Fig. \ref{fig:fig3}(c)) due to the limited range of variation of $C_Q$ in conjunction with the dissipation in the system.
	
	To interpret the experimental results of parametric behaviour and explore the design considerations for exploiting this effect in a practical QDPA, we develop a semi-classical model incorporating the small signal equivalent circuit of the QD device \cite{esterliSmallsignalEquivalentCircuit2019}. The fundamental mechanism of parametric amplification can be illustrated by considering the simplified circuit of Fig. \ref{fig:fig1}(a): a time-varying capacitor $C(t)$ connected to a band-limited load $Z(\omega)$ with an injected current $i_{in} = I_s\cos(\omega_s)$ as the input signal. Pumping $C(t) = \tilde{p} + \tilde{p}^*$, where $\tilde{p} = P e^{j\omega_pt}$, and considering the signal and idler mode voltage components within the passband of $Z(\omega)$, $v(t) = \tilde{v}_s + \tilde{v}_s^* + \tilde{v}_i + \tilde{v}_i^*$, parametric mixing produces a current $i_C = \frac{d}{dt}(CV)$ with frequency components $I_C(\omega_s) = j\omega_s \tilde{p}\tilde{v}_i^*$ and $I_C(-\omega_i)=-j\omega_i \tilde{p}^*\tilde{v}_s$. Pumping therefore converts energy from the signal to the idler mode ($\tilde{v}_i^* = j\omega_i \tilde{p}^*\tilde{v}_sZ^*(\omega_i)$) and vice versa ($\tilde{v}_s = (\tilde{i}_{in}-j\omega_s \tilde{p}\tilde{v}_s^*)Z(\omega_s)$), resulting in complex mode amplitudes:
	\begin{equation}
		\begin{split}
			V_s &= \frac{I_{in}Z(\omega_s)}{1-\omega_s\omega_i\left|P\right|^2Z(\omega_s)Z^*(\omega_i)}\\
			V_i^* &= \frac{j\omega_iP^*I_{in}Z(\omega_s)Z^*(\omega_i)}{1-\omega_s\omega_i\left|P\right|^2Z(\omega_s)Z^*(\omega_i)}
		\end{split}
		\label{eqn:gain}
	\end{equation}

	These expressions highlight the essential behaviour of the amplifier: gain is achieved as the denominator approaches zero at a critical pump amplitude \cite{mehrpooCryogenicCMOSParametric2020}, above which self-oscillation occurs. The frequency response is set by $Z(\omega)$, though the apparent 3~dB bandwidth decreases with pump power as the peak gain increases. The phase of the signal mode is independent of the phase of the pump, but the phase of the idler mode tracks the phase of $P$; in the degenerate case where $\omega_i = \omega_s$, interference between the signal and idler modes leads to a phase dependent gain. Applying this analysis to the full equivalent circuit incorporating the inductively coupled resonator of figure 1(c), we assimilate the $V_f$,$V_b$ and $V_r$ wave amplitudes at $\omega_s$ and $\omega_i$ into rank-4 matrix equation $\mathbf{A}\mathbf{V}_{r,b} = \mathbf{B}V_f + \mathbf{Z_n}\mathbf{I_n}$, where the gain matrices $\mathbf{A}(\omega_s, \omega_i, P)$ and $\mathbf{B}(\omega_s)$ and noise impedance $\mathbf{Z_n}(\omega_s, \omega_i)$ are functions of the resonator parameters $C, L_1, L_2, M$ and $R$ (see supplementary material). Technical noise sources such as thermal noise arising from dissipative elements of the resonator and Sisyphus cyclostationary noise associated with stochastic dot-reservoir tunnelling can be modelled by a noise current generator with spectral density ${I_n}(\omega)^2$ in parallel with the $LC$ resonator. 
	
\begin{figure}
		\includegraphics{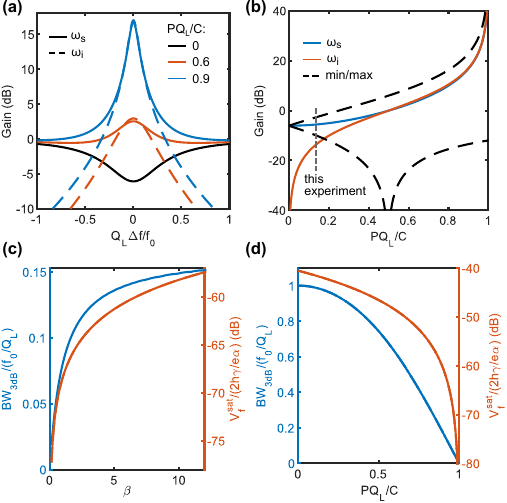}
		\caption{Simulated performance of the degenerate QDPA ($\omega_p = 4\pi f_0$) for $Q_i = 1500, Z_r = \SI{500}{\ohm}$ and $\beta = 3$. (a)  Frequency response of the signal (solid line) and idler (dashed line) at different pump amplitudes. (b) Signal and idler mode amplitudes as a function of pump amplitude. Dashed lines correspond to the minimum and maximum phase-dependent dependent gains. (c),(d) Dependence of normalised bandwidth and input saturation power on $\beta$ for pump amplitudes achieving a maximum gain of 20~dB (c) and pump amplitude for $\beta = 3$ (d). $V_f^\mathrm{sat}$ is the input amplitude that gives a gate voltage $V_{rs} + V_{ri}$ equal to the DTR transition FWHM $V_{1/2} = 2h\gamma/e\alpha$.
		\label{fig:fig4}}
	\end{figure}

	Figure \ref{fig:fig4}a summarises solutions of the matrix equation for a degenerate pumping scheme ($\omega_p = 4\pi f_0$), normalised in terms of the resonator design parameters ($f_0$, $Z_r$, $Q_L$ and $\beta$) and the `QD varactor' parameters ($\alpha$ and $\gamma$): we see that as the pump amplitude increases, the system produces a net gain around the natural frequency of the resonator. On resonance, the transmission line characteristic impedance $Z_0$ is transformed through the inductive coupling to a value $R_\mathrm{eff} = Q_e/\omega C$, giving a total impedance loading the varactor of $(1/R + 1/R_\mathrm{eff})^{-1} = Q_L/\omega C$. By analogy to eq. \ref{eqn:gain}, self-oscillations occur above a critical pump amplitude of $P_\mathrm{0} = C/Q_L$, with the gain increasing as $1/(1-P^2/P_\mathrm{0}^2)$; the minimum and maximum phase-dependent gains due to signal and idler mode interference are given by the sum and difference of the mode amplitudes (Fig. \ref{fig:fig4}(b)). Fig. \ref{fig:fig4}(c) shows a strongly overcoupled resonator design ($\beta \gg 1$) improves both the bandwidth and input saturation power at a pump amplitude that achieves a gain of 20~dB. As $P/P_0$ approaches 1, the normalised amplifier bandwidth and input saturation amplitude $V_f^\mathrm{sat}$ decrease (Fig. \ref{fig:fig4}(d)).
	
	This analysis suggests how a practical QDPA can be designed to maximise the bandwidth for a target gain by optimising the QD varactor in conjunction with the resonator properties. Beyond maximising the QD-gate lever arm $\alpha_\mathrm{R}$ to increase $C_Q$, small $f/\gamma$ is required to reduce the Sisyphus dissipation; however, as $C_Q \propto 1/\gamma$ when  $h\gamma \gtrsim kT_e$, this leads to a trade-off between dissipation and pump amplitude at high operating frequencies. The available pump modulation $P \lesssim C_Q/4$ \cite{derakhshanmamanChargeNoiseOverdrive2020} is most efficiently utilised by maximising the resonator impedance $Z_r$ and setting $Q_L = C/P$, enabling the critical power to be achieved while maximising the bandwidth. Given the target $Q_L$, maximising $Q_i$ to achieve the strongly overcoupled regime further optimises the bandwidth and dynamic range. 
	
	In the current experiment, we find $PQ_L/C \approx 0.15$ (see fig.\ref{fig:fig4}(b)), indicating increases in the resonator $Q_i$ would be necessary to achieve practical gains. Based on reasonable estimates of current technological capabilities ($Q_i = 2000$ for integrated resonators \cite{zhengRapidGatebasedSpin2019}), a bandwidth of 0.6 MHz at 1.8 GHz could be achieved for a gain of 15 dB. Several avenues for improving the performance could be explored. More complex impedance matching schemes, such as double-tuned transformers,  break the link between $R_\mathrm{eff}$ and $Q_L$, enabling larger bandwidths for the same gain \cite{mehrpooCryogenicCMOSParametric2020, mutusStrongEnvironmentalCoupling2014}. Using multiple quantum dots in parallel, such as in a linear array~\cite{hutinGateReflectometryProbing2019}, would increase the available pump modulation and allow operation at lower $Q_L$, resulting in a larger bandwidth. 
	
	In conclusion, we have proposed and demonstrated parametric amplification using quantum capacitance as an alternative non-linear element to Josephson junctions. In our CMOS QD device, the experimentally achievable gain was limited by the degradation of resonator internal quality factor by Sisyphus dissipation. However, based on a semi-classical circuit model, we expect design optimisations could enable performance comparable to JPAs.

	\begin{acknowledgments}
		This research has received funding from the European Union’s Horizon 2020 Research and Innovation Programme under Grant Agreement No. 688539 \footnote{http://mos-quito.eu.} L.C. acknowledges support from EPSRC Cambridge
		UP-CDT EP/L016567/1, T.L. acknowledges support from	EPSRC Cambridge NanoDTC EP/L015978/1 and M.F.G.Z. acknowledges support from the European Union’s grant agreement No. 951852, Innovate UK Industry Strategy Challenge Fund (10000965) and the UKRI Future Leaders Fellowship Programme (MR/V023284/1).
	\end{acknowledgments}

\end{document}